\documentclass[runningheads]{llncs}
\usepackage{graphicx}
\usepackage{subfigure}
\usepackage{algorithm}
\usepackage{algpseudocode}
\usepackage{amsmath}

\usepackage{booktabs}
\usepackage{multirow}
\usepackage{makecell}
\usepackage{float}
\usepackage{enumitem}
\usepackage[misc]{ifsym}
\usepackage{orcidlink}

\usepackage{enumitem}
\setlist[itemize]{topsep=2pt,parsep=2pt,itemsep=2pt,partopsep=2pt}
\setlist[itemize]{leftmargin=0pt}

\begin{document}
\title{PhishIntentionLLM: Uncovering Phishing Website Intentions through Multi-Agent Retrieval-Augmented Generation}
\titlerunning{PhishIntentionLLM - Accepted by ICDF2C 2025}
%
\author{WENHAO LI\inst{1}~\orcidlink{0009-0007-4342-6676},
SELVAKUMAR MANICKAM\inst{1}\textsuperscript{\Letter}~\orcidlink{0000-0003-4378-1954},
YUNG-WEY CHONG\inst{2}~\orcidlink{0000-0003-1750-7441} and
SHANKAR KARUPPAYAH\inst{1}~\orcidlink{0000-0003-4801-6370}}
\authorrunning{W. Li et al.}
%

\institute{Cybersecurity Research Centre, Universiti Sains Malaysia, Pulau Pinang, Malaysia \and
School of Computer Sciences, Universiti Sains Malaysia, Pulau Pinang, Malaysia\\
\email{wenhaoli@ieee.org, \{selva, chong, kshankar\}@usm.my}}
\maketitle              

\begin{abstract}
Phishing websites remain a major cybersecurity threat, yet existing methods primarily focus on detection, while the recognition of underlying malicious intentions remains largely unexplored. To address this gap, we propose \textit{PhishIntentionLLM}, a multi-agent retrieval-augmented generation (RAG) framework that uncovers phishing intentions from website screenshots. Leveraging the visual-language capabilities of large language models (LLMs), our framework identifies four key phishing objectives: Credential Theft, Financial Fraud, Malware Distribution, and Personal Information Harvesting. We construct and release the first phishing intention ground truth dataset (\textasciitilde2K samples) and evaluate the framework using four commercial LLMs. Experimental results show that \textit{PhishIntentionLLM} achieves a micro-precision of 0.7895 with GPT-4o and significantly outperforms the single-agent baseline with a \textasciitilde95\% improvement in micro-precision. Compared to the previous work, it achieves 0.8545 precision for credential theft, marking a \textasciitilde4\% improvement. Additionally, we generate a larger dataset of \textasciitilde9K samples for large-scale phishing intention profiling across sectors. This work provides a scalable and interpretable solution for intention-aware phishing analysis.

\end{abstract}
\section{Introduction}

Phishing is a dominant form of cybercrime that exploits both system vulnerabilities and human psychology to deceive users into disclosing sensitive information \cite{10.3389/fcomp.2021.563060,6497928}. Among its various forms, phishing websites pose one of the most critical threats, using visually deceptive interfaces to impersonate trusted entities \cite{10.1145/3589334.3645535}. The prevalence of such attacks has continued to grow, with the Anti-Phishing Working Group (APWG) reporting 989,123 unique phishing websites in the fourth quarter of 2024, compared to 888,585 in the same period of 2021 \cite{apwg}. This rise is fueled by phishing-as-a-service platforms, phishing toolkits, and affordable infrastructure as well as advanced evasive techniques phishers used \cite{tiis:101917,10175532}, making it easier for attackers to launch more long-lasting large-scale phishing campaigns. The impacts of these attacks go far beyond financial losses, including intellectual property theft and significant reputational damage \cite{10049452}.

To counter phishing website, a large body of research has focused on phishing website detection using heuristic-based, machine learning, and deep learning techniques that analyze features such as URLs, HTML structures, and domain metadata \cite{10788671,10633701}. Although these studies effectively detect phishing websites, none have focused specifically on identifying the malicious intentions behind these phishing websites.

\begin{figure*}[!ht]
	\centering  
	\subfigtopskip=0pt 
	\subfigbottomskip=2pt
	\subfigcapskip=1pt 
	\subfigure[Credentials Theft]{
		\label{dhl_ct}
		\includegraphics[width=0.42\linewidth, height=3.5cm]{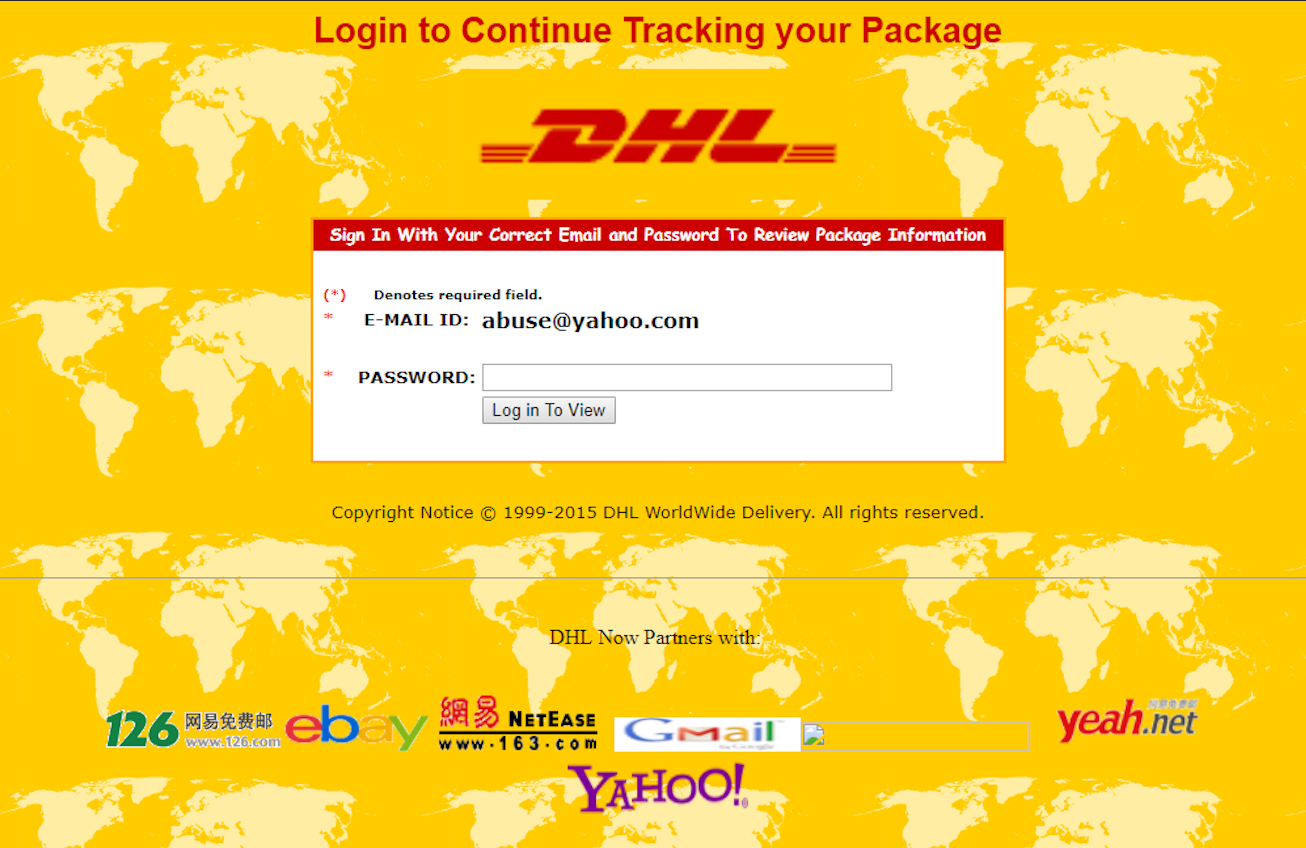}}
	\quad
	\subfigure[Malware Distribution]{
		\label{dhl_malware}
		\includegraphics[width=0.42\linewidth, height=3.5cm]{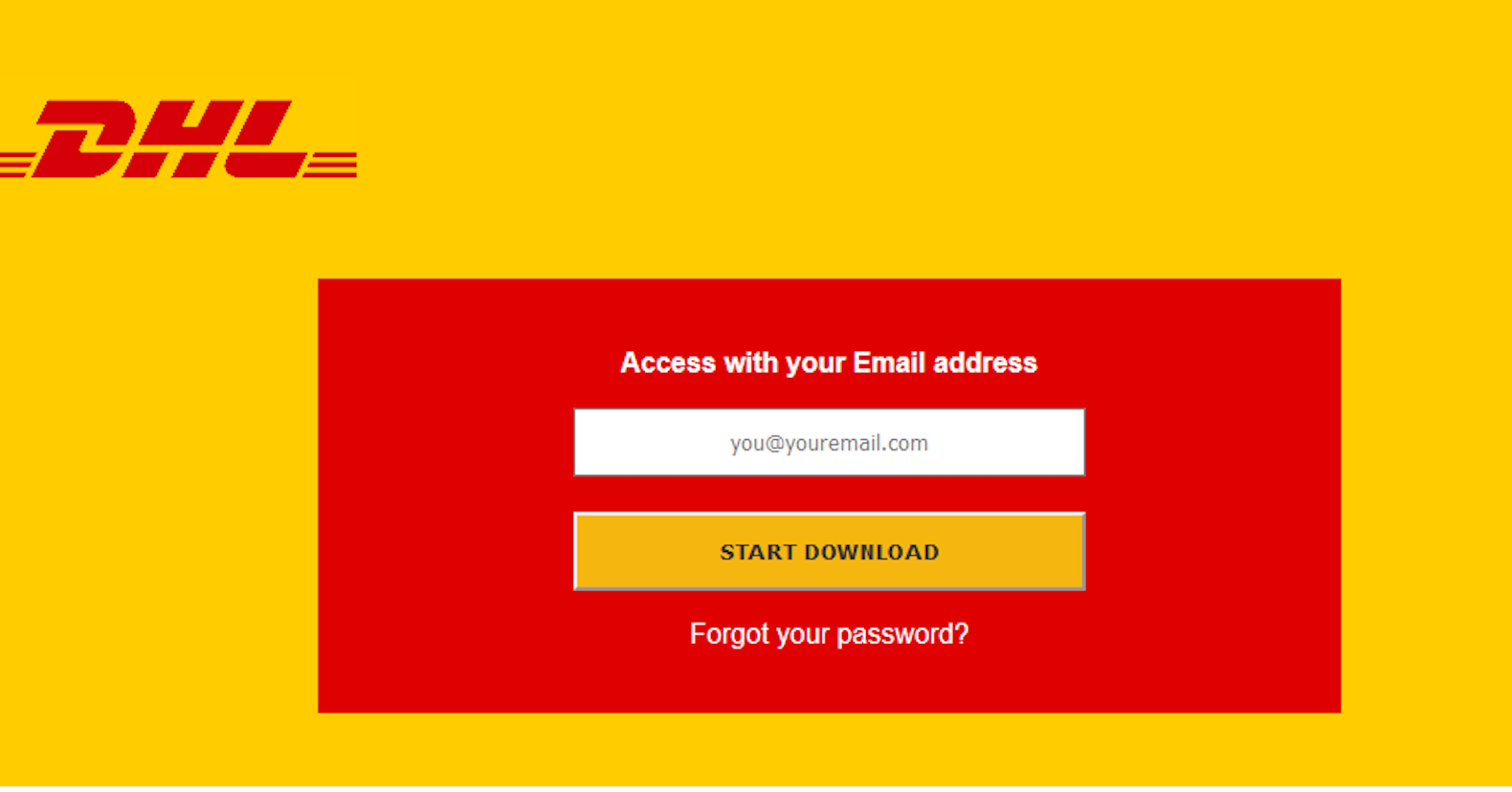}}
	\subfigure[Financial Fraud]{
		\label{dhl_financial}
		\includegraphics[width=0.42\linewidth, height=3.5cm]{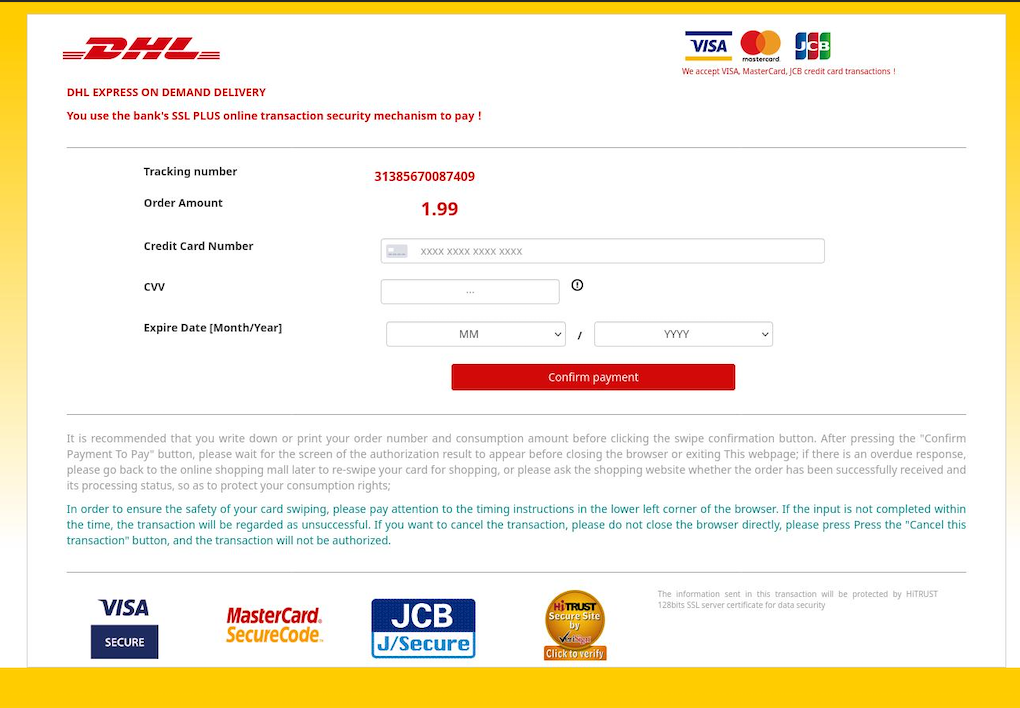}}
	\quad
	\subfigure[Personal Information Harvesting]{
		\label{dhl_personal}
		\includegraphics[width=0.42\linewidth, height=3.5cm]{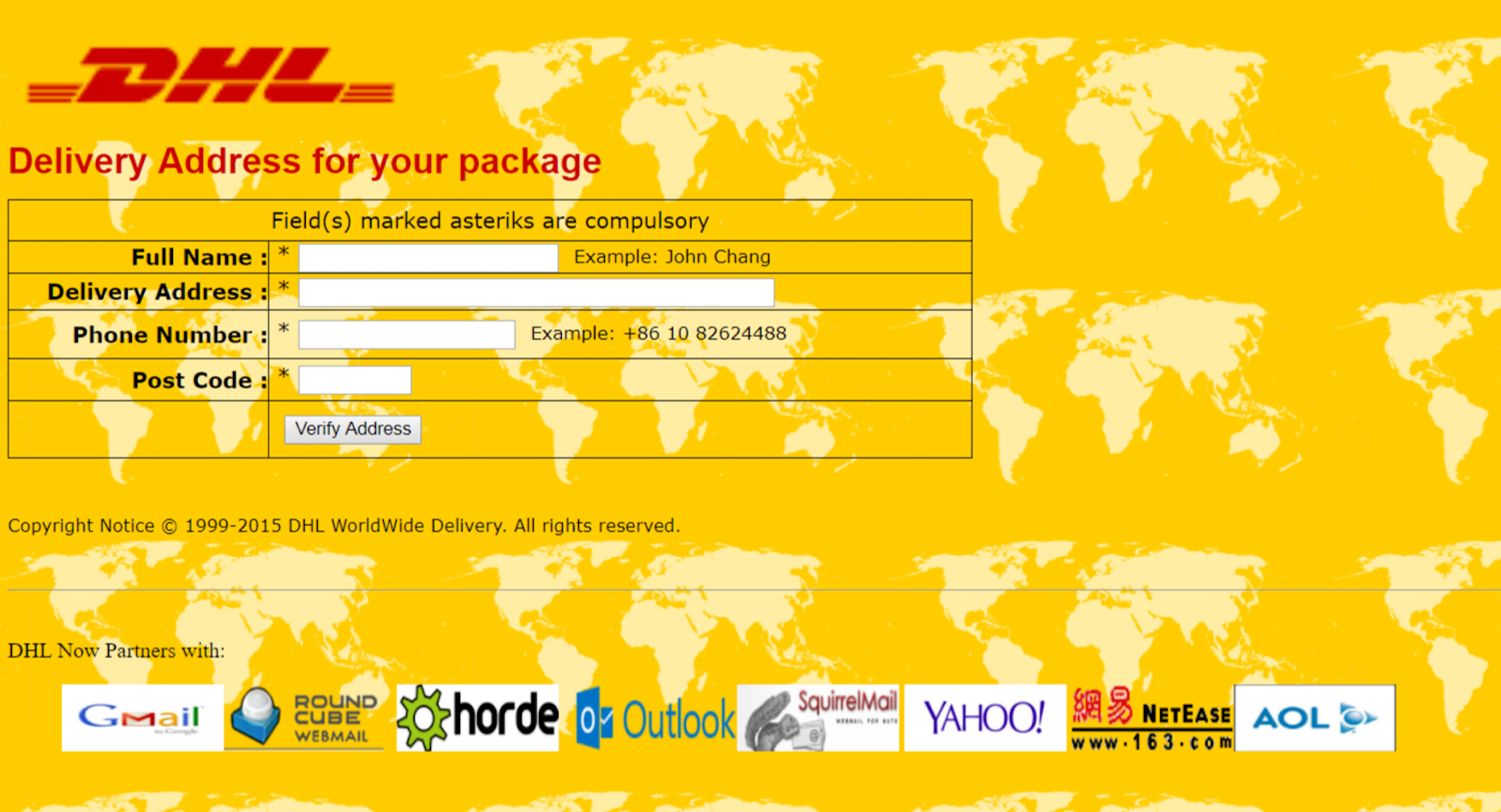}}
	\caption{The Phishing DHL Websites with varied Intentions.}
	\label{intentions}
\end{figure*}

Understanding the underlying intentions behind phishing websites provides deeper insights into attacker strategies and motivations. For instance, a phishing site targeting the banking sector with a credential theft intent presents a different threat profile than one designed to harvest healthcare data. Even phishing websites impersonating the same brand can exhibit diverse malicious goals, as illustrated in Fig.~\ref{intentions}. This level of granularity enables more precise threat intelligence, tailored defense strategies, and informed regulatory responses.

To bridge this gap, we propose a novel multi-agent retrieval-augmented generation (RAG) framework for identifying malicious intentions behind phishing websites through visual screenshot analysis. Our approach leverages the visual-language capabilities of large language models (LLMs) in combination with a domain-specific retrieval module to detect four primary phishing threat categories: Credential Theft, Financial Fraud, Malware Distribution, and Personal Information Harvesting.

The key contributions of this paper are as follows:
\begin{itemize}
    \item We manually construct and publicly release the first phishing intention ground truth dataset containing \textasciitilde 2K phishing website samples with screenshots \footnotemark[1], labeled intention categories, and sectoral information.
    \item We propose \textit{PhishIntentionLLM}, a novel multi-agent RAG framework that synergizes general and expert agents with knowledge bases for phishing intention detection.
    \item We comprehensively evaluate our framework using four commercial LLMs across multiple performance metrics, and benchmark it against a single-agent baseline and prior work focused on credential theft detection.
    \item We leverage our framework to detect the intention of a larger-scale phishing samples using GPT-4o and generate and publicly release a \textasciitilde 9K phishing intention dataset \footnotemark[1], analyzing the distribution of phishing intentions and their associated sectors to uncover empirical patterns in attacker behaviors.
    
\end{itemize}
\footnotetext[1]{Dataset: \url{https://github.com/v1ct0rl33/PhishIntentionLLM}}

The remainder of this paper is organized as follows: Section~\ref{sec:related_work} reviews related work. Section~\ref{sec:background} introduces four primary phishing intentions. Section~\ref{sec:method} describes the architecture of the proposed multi-agent framework. Section~\ref{sec:evaluation} outlines the models and metrics used for evaluation as well as the ground truth dataset construction. Section~\ref{sec:results} presents experimental results and comparative analysis. Section~\ref{sec:discussion} discusses a larger-scale empirical phishing intention analysis with proposed framework, while Section~\ref{sec:conclusion} concludes the paper with future works.

\section{Related Work}\label{sec:related_work}

Phishing detection has been widely studied using heuristics \cite{6825621}, machine learning and deep learning approaches that analyze features such as URLs \cite{SAHINGOZ2019345}, HTML content \cite{9207707}, domain metadata \cite{10.1145/3205977.3205992} or hybrid features \cite{ALJOFEY2025104170}. While these methods have proven effective in identifying phishing websites, many of these techniques rely on code-level or metadata features that can be easily obfuscated or manipulated, making them particularly vulnerable to cloaking techniques that hide malicious content from automated scanners while displaying convincing visuals to human victims \cite{10175532}.

To address limitations of feature-based methods, recent studies have explored visual and multimodal analysis \cite{272200}, using screenshots, layout structure, and Optical Character Recognition (OCR)-extracted text for phishing detection in addition to existing approaches \cite{liu2024cnn,10723311}. These approaches better mimic how human users perceive websites and are naturally resilient to cloaking techniques that hide malicious content in code while keeping visual appearance intact. By focusing on visual elements, such models can generalize across phishing pages that differ at the source-code level but share deceptive front-end appearances.

However, they offer limited insights into the nature or intent of the attack. Once a phishing website is detected, understanding its specific malicious goal (e.g., credential theft vs. malware distribution) is often overlooked, leaving a critical gap in threat profiling and response strategies.

Understanding attackers' intentions has gained attention in broader cybersecurity contexts especially in the field of cybercrime \cite{info15050263}, such as classifying types of email phishing (e.g., business email compromise vs. generic scams) \cite{10.1002/spy2.165}, ransomware behavior \cite{10917287}, and scams and fraud campaign strategies \cite{10.1145/3700838.3703672}. These studies demonstrate that identifying intent is crucial for targeted mitigation, forensic analysis, and policy-making. However, such intent-focused analysis has rarely been applied to phishing websites, especially in a structured, automated manner.

To date, only one known study has attempted to detect phishing website intentions, focusing specifically on credential theft \cite{279900}. While this work represents an important step forward and demonstrates that identifying intent can aid in detecting previously unknown phishing websites, it is limited to credential theft and does not address other potential malicious phishing intention. Therefore, a more scalable and generalizable approach is required to capture the full spectrum of phishing intentions.

Recent advancements in LLMs and RAG have enabled more context-aware and interpretable solutions across various natural language processing and security tasks \cite{osti_2474934,10847404}. These models integrate external knowledge retrieval with language understanding, offering robust performance on complex, multi-stage tasks.

In summary, although phishing detection has advanced through traditional and deep learning methods, most approaches focus solely on binary classification, neglecting the identification of underlying malicious objectives. While visual and multimodal analysis helps address obfuscation, it has yet to be applied to phishing intention detection. Existing intent recognition efforts in cybersecurity are limited, with only one study on phishing intent that targets credential theft and lacks support for multiple intentions. Moreover, powerful tools like RAG and LLMs remain underexplored in this context. This highlights the need for a comprehensive, scalable framework that leverages visual and contextual cues to identify diverse phishing intentions.

\section{Background}\label{sec:background}

Phishing websites employ various deceptive strategies, each designed with specific malicious intentions. As illustrated in Fig.~\ref{intentions}, our analysis of real-world phishing campaigns reveals four predominant categories of malicious intent: credential theft, malware distribution, financial fraud, and personal information harvesting. These intentions represent the primary objectives that drive phishing attacks in the current threat landscape. This section provides essential background on each category to establish a foundation for understanding the methodology presented in this study.

Credential theft, as shown in the Fig.\ref{dhl_ct}, perhaps the most common phishing objective, involves attackers creating counterfeit websites that mimic legitimate platforms to capture user authentication credentials. These attacks typically impersonate trusted entities such as financial institutions, email providers, or corporate platforms, employing visual and structural similarities to the original sites. Once obtained, these credentials facilitate account takeovers, enabling attackers to access sensitive information, conduct unauthorized transactions, or establish footholds for further attacks.

Malware distribution phishing leverages deceptive interfaces to induce users to download malicious software, as shown in the Fig.~\ref{dhl_malware}. Such attacks frequently masquerade as software updates, security scans, media players, or document viewers. The distributed malware may include ransomware, information stealers, remote access trojans, or other malicious payloads that compromise system integrity and user privacy. These attacks typically feature prominent download buttons, alarming security warnings, or counterfeit system notifications.

Financial fraud phishing specifically targets monetary exploitation through various schemes designed to manipulate victims into financial transactions, as shown in the Fig.~\ref{dhl_financial}. These attacks employ deceptive narratives including fake investment opportunities, fraudulent merchandise sales, technical support scams, and counterfeit financial alerts. Distinguished by their emphasis on payment information collection or direct transfer solicitation, these attacks often create artificial urgency to circumvent rational decision-making processes.

Personal information harvesting, as shown in the Fig.~\ref{dhl_personal}, aims to collect comprehensive personally identifiable information beyond mere credentials. These attacks solicit sensitive data including government identification numbers, home addresses, employment details, financial information, and healthcare data, often through illegitimate forms, surveys, or registration pages. This information enables identity theft, sophisticated social engineering, or sale on underground markets for subsequent exploitation.

\section{Methodology}\label{sec:method}
This section presents the proposed methodology of  this study. We describe the system's hierarchical agent architecture, knowledge retrieval mechanisms, processing pipeline. The formalized algorithm for this framework is proposed to demonstrate how these components interact to produce accurate threat classifications with supporting evidence chains, addressing the challenging task of multi-category phishing intention identification.

\subsection{System Architecture}
\textit{PhishIntentionLLM} represents a novel multi-agent RAG framework for identifying malicious intentions behind phishing websites through screenshot analysis. Our approach leverages the visual understanding capabilities of LLMs combined with specialized RAG to detect up to four primary threat categories: Credential Theft, Financial Fraud, Malware Distribution, and Personal Information Harvesting.

\begin{figure}[!ht]
\centering
\includegraphics[width=0.90\linewidth]{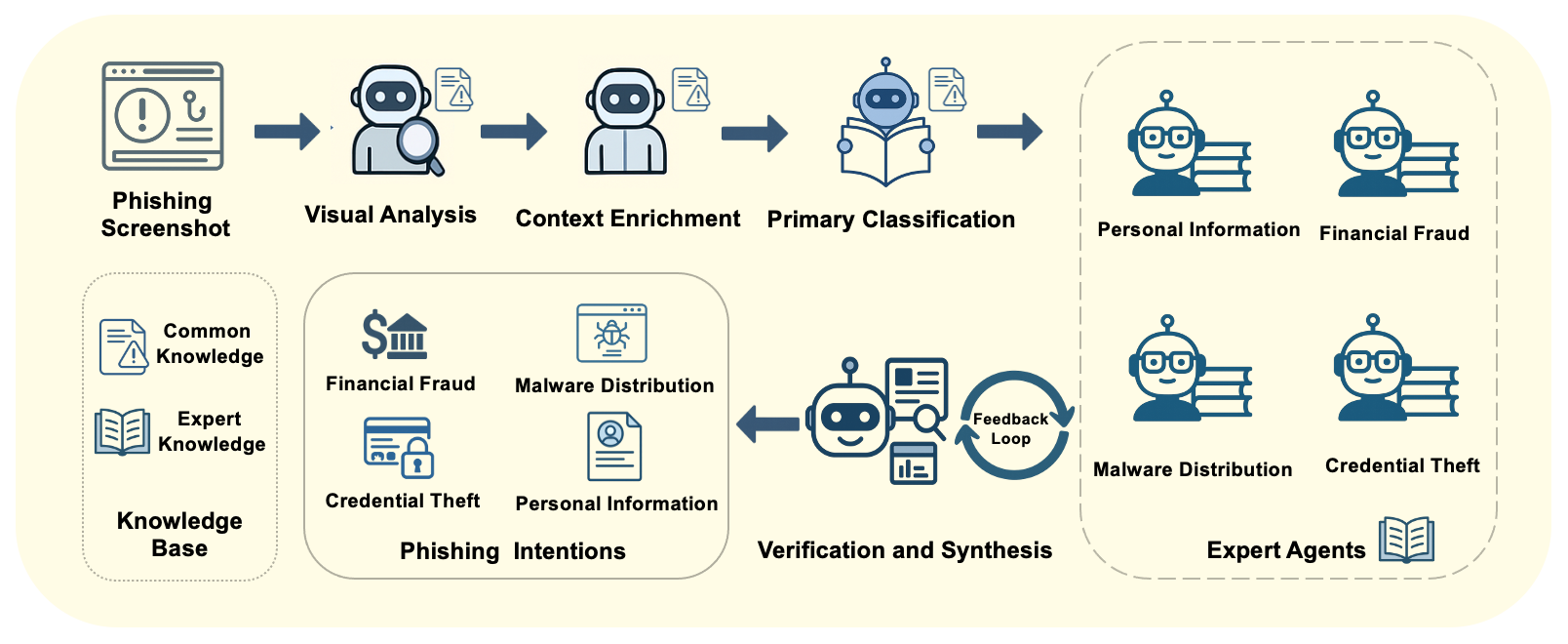}
\caption{Proposed PhishIntentionLLM Framework.}
\label{method}
\end{figure}

The Fig.~\ref{method} presents an overview of the proposed framework. The system employs a hierarchical multi-agent architecture comprising five specialized layers, each with distinct cognitive responsibilities, working in conjunction with domain-specific knowledge bases. The Vision Analysis Agent serves as the perception layer, responsible for extracting raw data from phishing website screenshots. Using vision-language models, it identifies and organizes visual elements including textual content, interface components, page layout, and domain information when available.

The Context Enrichment Agent functions as a semantic layer that enhances extracted elements with security-relevant context. This agent retrieves basic threat patterns from the knowledge base, tags suspicious elements with contextual information, maps visual elements to potential security implications, and generates preliminary threat hypotheses based on established patterns in phishing detection.

The Classification Agent performs multi-label classification to identify the most likely threat categories. It calculates confidence scores for each threat category, selects one to three primary threats for deeper analysis, associates evidence with each identified type, and creates an initial classification hypothesis that guides subsequent specialist analysis.
The Specialist Analysis Layer contains four expert agents, each dedicated to a specific threat category. When activated based on initial classification, these specialists conduct in-depth analysis within their domains. The Credential Theft Agent evaluates login form characteristics and domain spoofing patterns, while the Financial Fraud Agent analyzes payment solicitation and unrealistic financial promises. Similarly, the Malware Distribution Agent examines download prompts and software update impersonation, and the Personal Information Agent assesses excessive data collection and privacy policy issues.

The Validation Agent integrates all previous analyses to form a comprehensive assessment. This agent combines specialist findings, resolves potential conflicts between competing hypotheses, weighs evidence based on reliability and relevance, and produces a final classification with complete evidence chains and associated confidence scores.

\subsection{Knowledge Retrieval Architecture}

\textit{PhishIntentionLLM} incorporates a dual-layer knowledge architecture, as shown in the Definition~\ref{defin}, that augments agent reasoning through retrieval-augmented generation. The Basic Threat Pattern Repository contains domain-agnostic phishing indicators, including common deception patterns, visual deception elements, suspicious text patterns, and URL red flags. This knowledge supports initial detection and context enrichment phases, providing foundational patterns that transcend specific threat categories.

The Category-Specific Knowledge Repository is organized by threat type and provides detailed domain knowledge. For Credential Theft, it includes common targets, obfuscation techniques, and form submission patterns. The Financial Fraud section contains scam typologies, pressure tactics, and payment anomalies. Malware Distribution knowledge encompasses malware types, download mechanisms, and system access requests, while Personal Information resources detail data collection patterns and privacy indicators. This specialized knowledge enables expert agents to conduct nuanced analyses within their respective domains.

\begin{definition}[Knowledge Base Structures]\label{defin}
The PhishIntentionLLM framework employs two complementary knowledge repositories:

\begin{enumerate}
    \item \textbf{Basic Threat Pattern Repository} ($K_B$):
    
    \begin{equation}
    K_B = \{P_c, P_v, P_t\}
    \end{equation}
    
    where:
    \begin{itemize}
        \item $P_c = \{p_1, p_2, \ldots, p_n\}$ is the set of common phishing patterns
        \item $P_v = \{v_1, v_2, \ldots, v_m\}$ is the set of visual deception techniques
        \item $P_t = \{t_1, t_2, \ldots, t_k\}$ is the set of text-based manipulation patterns
    \end{itemize}
    
    \item \textbf{Specialist Knowledge Repository} ($K_C$):
    
    For each threat category $c \in \{$Credential Theft, Financial Fraud, Malware Distribution, Personal Information Harvesting$\}$, we maintain:
    
    \begin{equation}
    K_C^c = \{F^c, D^c\}
    \end{equation}
    
    where:
    \begin{itemize}
        \item $F^c = \{f_1^c, f_2^c, \ldots, f_p^c\}$ is the set of primary features for category $c$
        \item $D^c = \{T^c, M^c, I^c\}$ is the detailed knowledge structure, containing:
        \begin{itemize}
            \item $T^c = \{t_1^c, t_2^c, \ldots, t_q^c\}$: common targets 
            \item $M^c = \{m_1^c, m_2^c, \ldots, m_r^c\}$: specialized techniques
            \item $I^c = \{i_1^c, i_2^c, \ldots, i_s^c\}$: distinctive indicators
        \end{itemize}
    \end{itemize}
\end{enumerate}
\end{definition}

\subsection{Processing Pipeline}
The \textit{PhishIntentionLLM} processing workflow begins with image input, where the system ingests website screenshots as primary data. The Visual Analysis phase extracts text, interface elements, and page structure using vision-language capabilities. During Context Enhancement, the extracted elements are enriched with security context from the basic threat repository, establishing preliminary security implications for observed elements.
Initial Classification identifies one to three primary threat categories and calculates confidence scores, determining which specialist agents to activate. In the Specialist Analysis phase, these activated agents perform in-depth examination within their respective threat domains, applying category-specific knowledge to the enriched context. Evidence Synthesis integrates all analyses to produce a cohesive final classification with complete evidence chains.
When confidence falls below established thresholds, a feedback loop activates additional specialist analyses to improve certainty. This adaptive mechanism allows the system to handle ambiguous cases by gathering additional perspectives. Finally, Result Generation outputs identified threats with confidence scores and supporting evidence, providing a comprehensive assessment of phishing intentions.

\begin{figure}[!ht]
\centering
\includegraphics[width=0.65\linewidth]{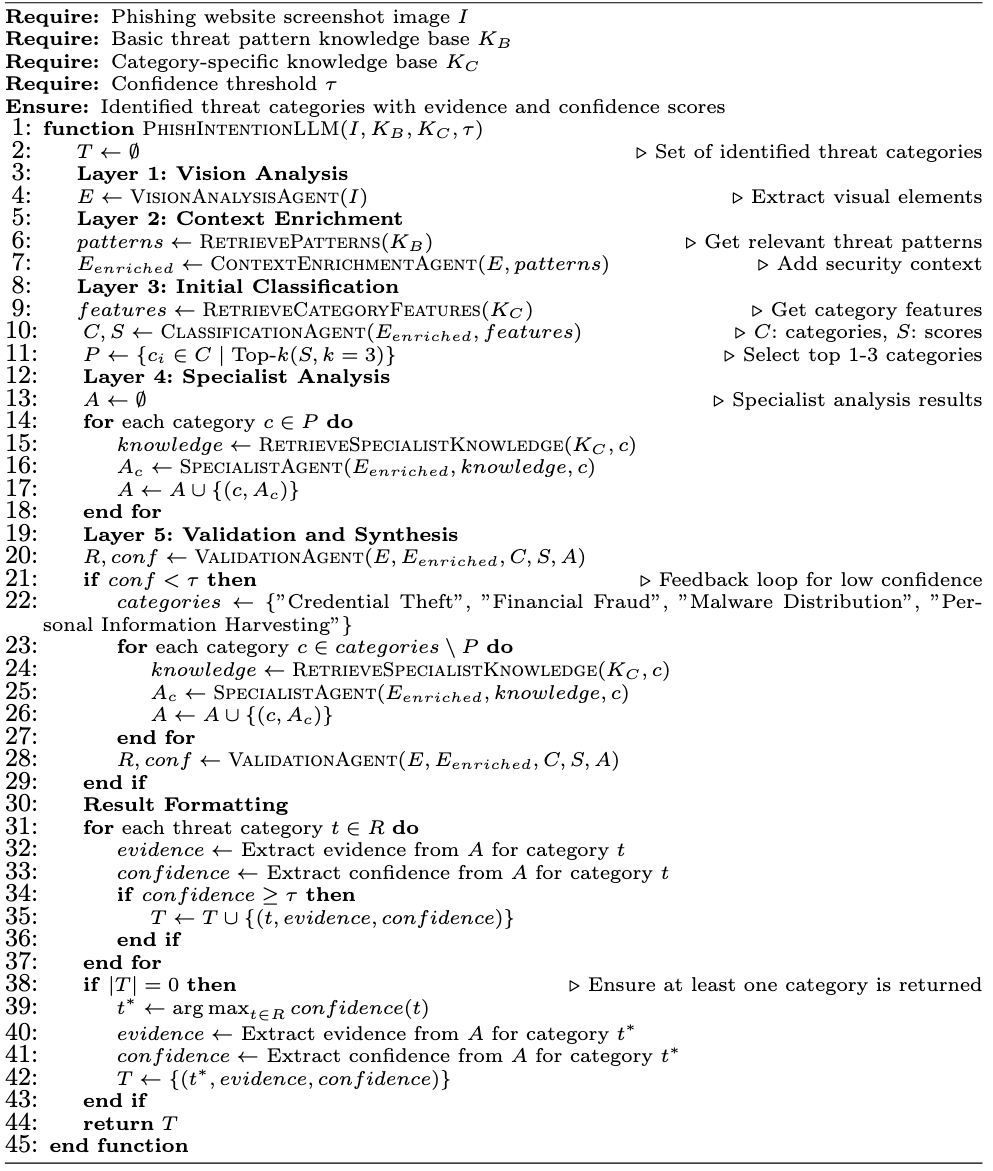}
\caption{The Proposed Algorithm for PhishIntentionLLM Framework.}
\label{alg:phishintentionllm}
\end{figure}

Fig.~\ref{alg:phishintentionllm} demonstrates the \textit{PhishIntentionLLM} approach through precise computational stages and decision procedures. The algorithm emphasizes the mathematical relationship between input screenshot analysis and threat categorization using set operations and conditional branches. Notable features include the top-k selection function for candidate threat categories (line 11), the union operation for specialist analysis aggregation (line 19), and the argmax function for determining the highest confidence category when necessary (line 39).

\section{Evaluation}\label{sec:evaluation}
\subsection{Model Selection}
To evaluate the effectiveness of our framework, we selected four state-of-the-art multimodal LLMs with diverse architectures and capabilities. Qwen2.5-VL-72B-Instruct \cite{bai2025qwen2} is a 72B-parameter vision-language model by Alibaba Cloud, pre-trained on multilingual and multimodal datasets and known for strong visual reasoning performance. Gemini-2.0-Flash-001 \cite{team2023gemini}, developed by Google DeepMind, balances efficient inference and advanced multimodal processing. GPT-4o \cite{gpt4o}, OpenAI’s flagship model, integrates high-level visual understanding with robust reasoning, while GPT-4o-mini \cite{gpt4omini} offers a more lightweight alternative that retains strong multimodal capabilities. All models support API integration, ensuring compatibility with our framework and enabling comparative evaluation across computational and architectural dimensions.

\subsection{Ground Truth Dataset}
To construct the ground truth dataset for the evaluation, we randomly selected phishing website samples from three existing datasets that contain screenshots for each phishing instance and manually removed the samples with poor quality of screenshots \cite{272200,putra_2023_8041387,8567299}. The labeling process involved three cybersecurity engineers, each with a minimum of three years of professional experience. Two engineers independently labeled the intentions of these phishing samples based on the screenshots, while the third engineer reviewed all labeled samples to ensure consistency and accuracy. The engineers assigned one or multiple intentions to each phishing sample based on their visual content analysis. 

For instance, if a phishing website solely requested username and password credentials, it was labeled as "Credential Theft." However, if the website additionally solicited telephone numbers or address information, it received dual labels of "Credential Theft" and "Personal Information Harvesting."

Through this rigorous labeling process, we constructed a multi-intention phishing dataset comprising 2,063 phishing samples from domains such as e-commerce, finance, social networking, telecommunications, and delivery services. This is one of the first phishing datasets to include explicit phishing intentions. The ground truth dataset has been publicly released.

\begin{table}[htbp]
\centering
\scriptsize 
\caption{Distribution of Phishing Intentions and Number of Intentions per Record.}
\label{tab:intention_counts}
\begin{tabular}{@{}llc@{}}
\toprule
\textbf{Category} & \textbf{Type} & \textbf{Count} \\
\midrule
\multirow{4}{*}{Phishing Intention Type} 
    & Credentials Theft & 1696 \\
    & Malware Distribution & 68 \\
    & Financial Fraud & 222 \\
    & Personal Information Harvesting & 408 \\
\midrule
\multirow{3}{*}{Number of Intentions per Record}
    & One Intention & 1757 \\
    & Two Intentions & 281 \\
    & Three Intentions & 25 \\
\bottomrule
\end{tabular}
\end{table}

Table~\ref{tab:intention_counts} presents the distribution of phishing intentions identified in ground truth dataset, along with the number of records containing one, two, or three distinct intentions. The majority of records are associated with Credentials Theft (1,696), followed by Personal Information Harvesting (408), Financial Fraud (222), and Malware Distribution (68). Additionally, most records contain only a single phishing intention (1,757), while fewer records exhibit two (281) or three (25) co-occurring intentions. No record contains four intentions.

\subsection{Evaluation Metrics}

To comprehensively evaluate this framework, various evaluation metrics are used in this study. The overall accuracy measures the proportion of correctly classified website screenshots:

\begin{equation}
\text{Accuracy} = \frac{|\{s \in S : \hat{Y}_s = Y_s\}|}{|S|}
\end{equation}

\noindent where $S$ represents all samples, $Y_s$ the true intention labels, and $\hat{Y}_s$ the predicted labels for sample $s$. 

Since phishing websites often exhibit multiple intentions simultaneously and some malicious phishing intentions (e.g., credential theft) are likely to occur more frequently than others (e.g., malware distribution) which leads to imbalanced classes, we employ micro-averaged metrics that aggregate contributions from all classes and reflect the actual data distribution, which is crucial in real-world scenarios:

\begin{equation}
\text{Precision}_{\text{micro}} = \frac{\sum_{c=1}^{C} TP_c}{\sum_{c=1}^{C} (TP_c + FP_c)}
\end{equation}

\begin{equation}
\text{Recall}_{\text{micro}} = \frac{\sum_{c=1}^{C} TP_c}{\sum_{c=1}^{C} (TP_c + FN_c)}
\end{equation}

\begin{equation}
\text{F1}_{\text{micro}} = \frac{2 \times \text{Precision}_{\text{micro}} \times \text{Recall}_{\text{micro}}}{\text{Precision}_{\text{micro}} + \text{Recall}_{\text{micro}}}
\end{equation}

\noindent where $C=4$ represents our intention classes, with $TP_c$, $FP_c$, and $FN_c$ denoting true positives, false positives, and false negatives for class $c$, respectively.

We also introduce Accuracy by Complexity ($\text{Acc}_{\text{comp}}$) to address the nuanced nature of multi-intention phishing websites:

\begin{equation}
\text{Acc}_{\text{comp}}(k) = \frac{|\{s \in S_k : \text{match}(Y_s, \hat{Y}_s) \geq t_k\}|}{|S_k|}
\end{equation}

\noindent where $S_k$ represents the set of samples with exactly $k$ intentions, $\text{match}(Y_s, \hat{Y}_s)$ counts the number of correctly matched intentions, and $t_k$ is the threshold for samples with $k$ intentions, defined as:

\begin{equation}
t_k = 
\begin{cases}
1, & \text{if } k = 1 \\
1, & \text{if } k = 2 \\
2, & \text{if } k = 3
\end{cases}
\end{equation}

This lies in the fact that for many phishing websites, even experienced cybersecurity experts can have slightly different understandings of the underlying intentions. Therefore, we believe partial matches provide valuable insights when evaluating multi-intention scenarios. We applied this to all above evaluations to ensure fair comparison against selected models.

For individual intention analysis, we calculate standard metrics for each class $c$ in our four phishing intention categories:

\begin{equation}
\text{Precision}_c = \frac{TP_c}{TP_c + FP_c} \quad
\text{Recall}_c = \frac{TP_c}{TP_c + FN_c}
\end{equation}

\begin{equation}
\text{F1}_c = \frac{2 \times \text{Precision}_c \times \text{Recall}_c}{\text{Precision}_c + \text{Recall}_c} \quad
\text{Accuracy}_c = \frac{TP_c + TN_c}{TP_c + TN_c + FP_c + FN_c}
\end{equation}

\noindent where $TN_c$ represents samples correctly identified as not belonging to class $c$.

\section{Results}\label{sec:results}
\subsection{Model Performance}
\begin{figure}[!ht]
\centering
\includegraphics[width=0.70\linewidth]{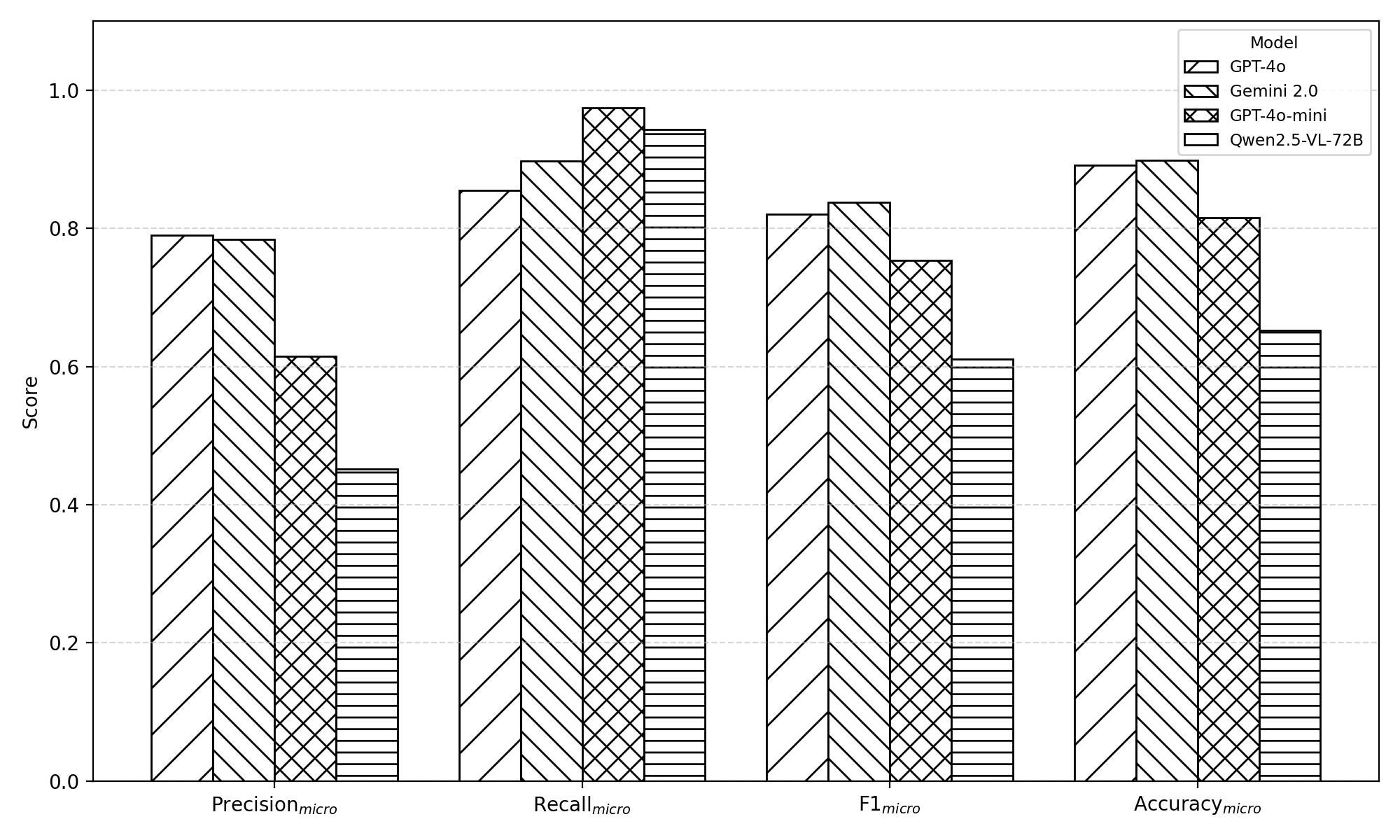}
\caption{Comparison of Selected LLMs on General Performance Metrics.}
\label{micro_metrics}
\end{figure}

Table~\ref{tab:llm_general_metrics} and Fig.~\ref{micro_metrics} present the general performance metrics of different LLMs integrated into our \textit{PhishIntentionLLM} framework. GPT-4o achieved the highest precision (0.7895) among all evaluated models, making it particularly valuable for phishing intention detection. High precision in this context means the model correctly identifies specific phishing intentions with minimal misclassifications, providing security analysts with more accurate understanding of attackers' objectives.

\begin{table}[htbp]
\centering
\scriptsize
\caption{General Performance Metrics of Selected LLMs with PhishIntentionLLM.}
\label{tab:llm_general_metrics}
\begin{tabular}{@{}lcccc@{}}
\toprule
\textbf{Model} & 
\textbf{Precision$_{micro}$} & 
\textbf{Recall$_{micro}$} & 
\textbf{F1$_{micro}$} & 
\textbf{Accuracy$_{micro}$} \\
\midrule
GPT-4o & 0.7895 & 0.8544 & 0.8207 & 0.8915 \\
Gemini 2.0 & 0.7843 & 0.8976 & 0.8371 & 0.8987 \\
GPT-4o-mini & 0.6149 & 0.9744 & 0.7540 & 0.8149 \\
Qwen2.5-VL-72B & 0.4520 & 0.9428 & 0.6111 & 0.6518 \\
\bottomrule
\end{tabular}
\end{table}

While GPT-4o excels in precision, it maintains strong performance across other metrics with a micro-averaged F1 score of 0.8207 and accuracy of 0.8915. This balanced performance demonstrates GPT-4o's effectiveness as a foundation model for identifying the underlying intentions of phishing campaigns. Gemini 2.0 follows closely with slightly lower precision (0.7843) but higher recall (0.8976), resulting in the highest overall F1 score (0.8371) and accuracy (0.8987). This indicates Gemini 2.0 identifies more actual phishing intentions at the expense of slightly more misattributed intentions compared to GPT-4o. GPT-4o-mini and Qwen2.5-VL-72B demonstrate significantly different performance characteristics, with extremely high recall values (0.9744 and 0.9428 respectively) but considerably lower precision scores (0.6149 and 0.4520). This suggests these models excel at capturing all potential intentions behind phishing websites but more frequently assign incorrect intentions, potentially complicating threat intelligence and response prioritization.

Table~\ref{tab:llm_complexity_accuracy} provides insights into model performance across different phishing intention complexity levels. Interestingly, the performance ranking shifts when evaluated using Accuracy by Complexity (Acc$_\text{comp}$) metrics.

\begin{table}[htbp]
\centering
\scriptsize
\caption{Comparison of Selected LLMs on (Acc$_\text{comp}$) and Overall Accuracy}
\label{tab:llm_complexity_accuracy}
\begin{tabular}{@{}lcccc@{}}
\toprule
\textbf{Model} & 
\makecell{\textbf{Acc$_\text{comp}$}\\\textbf{(1 Intention)}} & 
\makecell{\textbf{Acc$_\text{comp}$}\\\textbf{(2 Intentions)}} & 
\makecell{\textbf{Acc$_\text{comp}$}\\\textbf{(3 Intentions)}} & 
\makecell{\textbf{Overall}\\\textbf{Accuracy}} \\
\midrule
GPT-4o-mini & 0.8980 & 1.0000 & 0.9600 & 0.9130 \\
GPT-4o & 0.8996 & 0.9324 & 0.8800 & 0.9039 \\
Qwen2.5-VL-72B & 0.8953 & 0.9644 & 0.8800 & 0.9045 \\
Gemini 2.0 & 0.8750 & 0.9927 & 0.8800 & 0.8910 \\
\bottomrule
\end{tabular}
\end{table}

GPT-4o-mini demonstrated the highest overall accuracy (0.9130) across all complexity levels. The model achieved strong performance on single-intention phishing sites (0.8980) and notably high scores for multi-intention websites (1.0000 for two intentions and 0.9600 for three intentions).

GPT-4o maintains strong performance across complexity levels with high scores for single-intention phishing sites (0.8996), two-intention sites (0.9324), and three-intention scenarios (0.8800). This consistent performance further validates its reliability for accurately identifying diverse phishing strategies.

The Acc$_\text{comp}$ metrics for websites with three intentions show consistent performance across most models (0.8800), with only GPT-4o-mini achieving a higher score (0.9600). This suggests that the \textit{PhishIntentionLLM} framework provides reliable intention identification even in the most complex phishing scenarios with multiple malicious objectives.

These results demonstrate that while GPT-4o provides the most precise identification of specific phishing intentions, all evaluated models show strong capabilities in identifying phishing intentions across various complexity levels when integrated with our \textit{PhishIntentionLLM} framework.

\subsection{PhishIntentionLLM vs. Single-Agent}
\begin{figure}[!ht]
\centering
\includegraphics[width=0.70\linewidth]{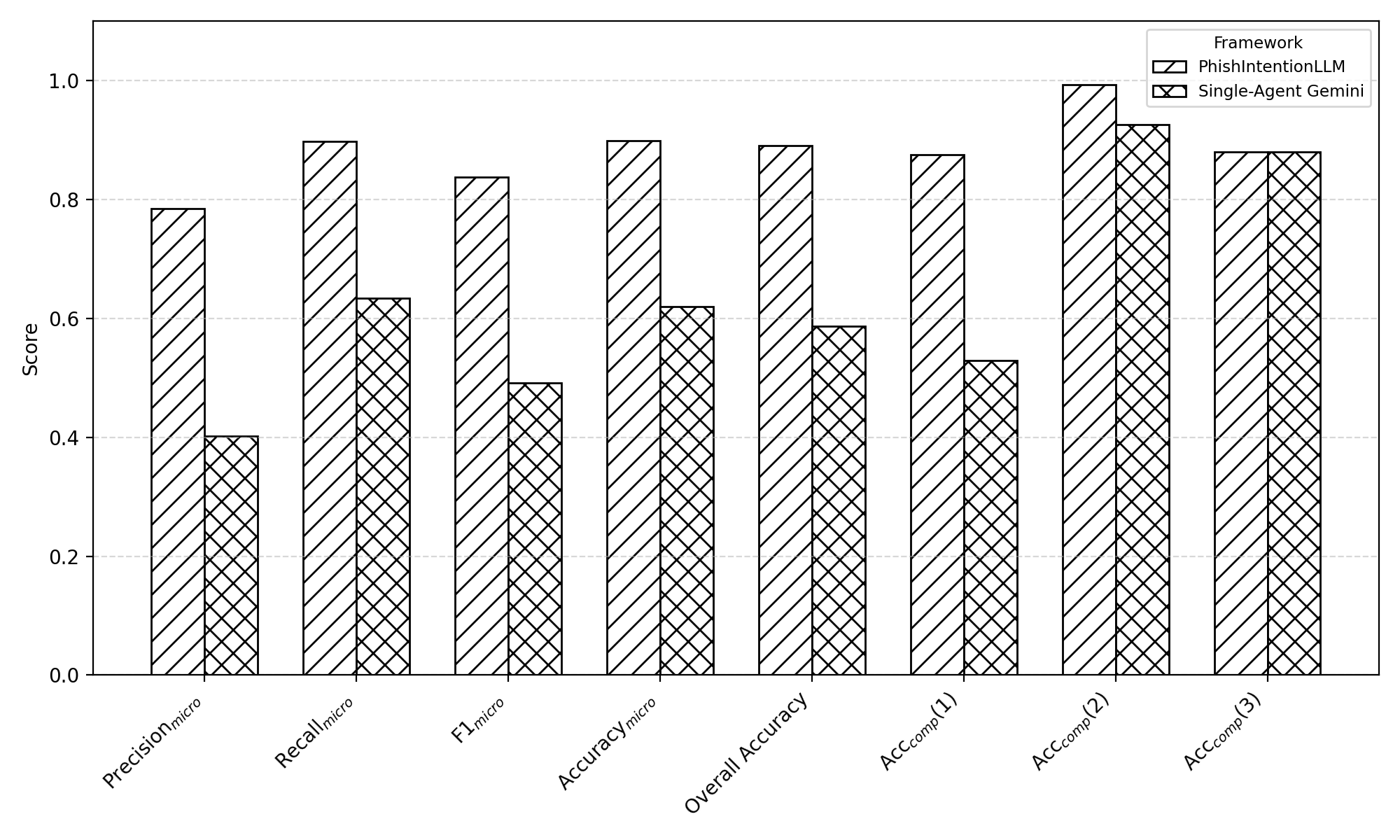}
\caption{Comparison of PhishIntentionLLM and Single-Agent Scenario.}
\label{agents_comparison}
\end{figure}

To further validate the effectiveness of our proposed framework, we conducted a comparative analysis between \textit{PhishIntentionLLM} and a single-agent baseline using identical foundation model (Gemini 2.0). As shown in Table~\ref{tab:multi_vs_single} and Fig.~\ref{agents_comparison}, the multi-agent RAG approach substantially outperforms the single-agent scenario across nearly all metrics.

\begin{table}[htbp]
\centering
\scriptsize
\caption{Comparison of PhishIntentionLLM and Single-Agent Scenario}
\label{tab:multi_vs_single}
\begin{tabular}{@{}l@{\hspace{15pt}}c@{\hspace{15pt}}c@{}}
\toprule
\textbf{Metric} & \makecell{\textbf{PhishIntentionLLM} \\ \textbf{(Gemini 2.0)}} & \makecell{\textbf{Single-Agent} \\ \textbf{(Gemini 2.0)}} \\
\midrule
Precision$_{micro}$ & \textbf{0.7843} & 0.4014 \\
Recall$_{micro}$ & \textbf{0.8976} & 0.6341 \\
F1$_{micro}$ & \textbf{0.8371} & 0.4916 \\
Accuracy$_{micro}$ & \textbf{0.8987} & 0.6195 \\
Overall Accuracy & \textbf{0.8910} & 0.5870 \\
Acc$_\text{comp}$(1 Intention) & \textbf{0.8750} & 0.5287 \\
Acc$_\text{comp}$(2 Intentions) & \textbf{0.9927} & 0.9253 \\
Acc$_\text{comp}$(3 Intentions) & 0.8800 & 0.8800 \\
\bottomrule
\end{tabular}
\end{table}

The most striking improvement appears in precision, where \textit{PhishIntentionLLM} achieves 0.7843 compared to just 0.4014 for the single-agent approach which represents a \textasciitilde 95\% improvement. This dramatic enhancement in precision demonstrates that our hierarchical agent architecture with specialized knowledge bases significantly reduces false positive classifications, enabling much more accurate identification of specific phishing intentions.

Similarly, recall improved from 0.6341 to 0.8976 (\textasciitilde{}42\% increase), indicating the multi-agent system's superior ability to identify all relevant intentions present in phishing websites. The combined improvements in precision and recall culminate in a \textasciitilde{}70\% increase in F1 score (0.8371 vs. 0.4916) and a \textasciitilde{}45\% enhancement in micro-accuracy (0.8987 vs. 0.6195).

The Accuracy by Complexity (Acc$_\text{comp}$) metrics reveal that our approach demonstrates particularly high effectiveness against phishing websites with single and two malicious intentions. For single-intention websites, \textit{PhishIntentionLLM} achieves 0.8750 accuracy compared to 0.5287 for the single-agent approach with a \textasciitilde{}65\% improvement.

\subsection{PhishIntentionLLM vs. PhishIntention}
To benchmark our framework against existing methods, we compared \textit{PhishIntentionLLM} (GPT-4o) with PhishIntention \cite{279900}, the only prior work focused on phishing intention analysis. Since PhishIntention is limited to detecting credential theft intentions only, we conducted a focused comparison on this specific intention type regarding test precision, test accuracy, F1 and recall, following PhishIntention's methodology with the same 9:1 (training:testing) ratio on our ground truth dataset.

\begin{figure}[!ht]
\centering
\includegraphics[width=0.70\linewidth]{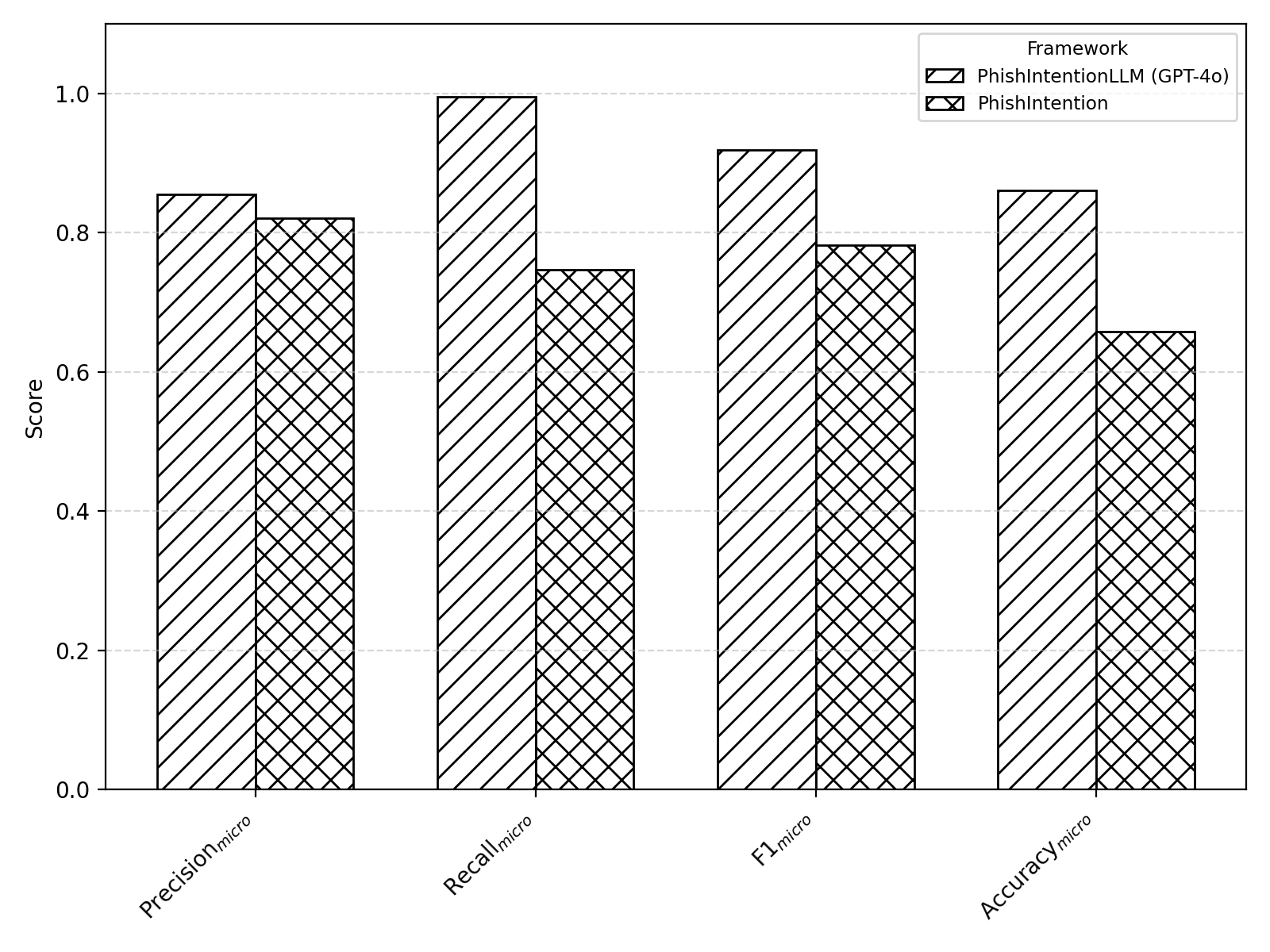}
\caption{Comparison of PhishIntention and PhishIntentionLLM on Credentials Theft Intention.}
\label{fig:phishintention_credentials}
\end{figure}

As shown in Table~\ref{tab:phishintention_credentials} and Fig.~\ref{fig:phishintention_credentials}, \textit{PhishIntentionLLM} demonstrates superior performance across all metrics in credential theft detection. Our framework achieved a precision of 0.8545 compared to PhishIntention's 0.8206, representing a \textasciitilde 4\% improvement in correctly identifying genuine credential theft attempts while reducing false positives.

\begin{table}[htbp]
\centering
\scriptsize
\caption{Comparison of PhishIntention and PhishIntentionLLM}
\label{tab:phishintention_credentials}
\begin{tabular}{@{}l@{\hspace{12pt}}c@{\hspace{12pt}}c@{\hspace{12pt}}c@{\hspace{12pt}}c@{}}
\toprule
\textbf{Model} & \textbf{Precision} & \textbf{Recall} & \textbf{F1 Score} & \textbf{Accuracy} \\
\midrule
PhishIntention \cite{279900} & 0.8206 & 0.7471 & 0.7821 & 0.6578 \\
\textbf{PhishIntentionLLM (Ours)} & \textbf{0.8545} & \textbf{0.9946} & \textbf{0.9193} & \textbf{0.8602} \\
\bottomrule
\end{tabular}
\end{table}

The most dramatic improvement appears in recall, where \textit{PhishIntentionLLM} achieved 0.9946 compared to PhishIntention's 0.7471 with a \textasciitilde 33\% increase. This substantial enhancement in recall indicates that our multi-agent RAG framework can identify virtually all credential theft attempts (99.5\%), whereas the existing approach misses approximately one-quarter of such phishing attempts.

The combined improvements in precision and recall result in a significant enhancement in F1 score (0.9193 vs. 0.7821), representing an \textasciitilde 18\% increase over the existing approach. Additionally, overall accuracy improved from 0.6578 to 0.8602, a \textasciitilde 31\% enhancement that demonstrates the superior classification capabilities of our framework. PhishIntentionLLM outperforms PhishIntention by fusing visual-text understanding with RAG-driven knowledge, unlike the latter’s static visual approach.

\section{Discussion}\label{sec:discussion}
\begin{figure}[!ht]
\centering

\subfigure[]{%
  \includegraphics[width=0.40\linewidth]{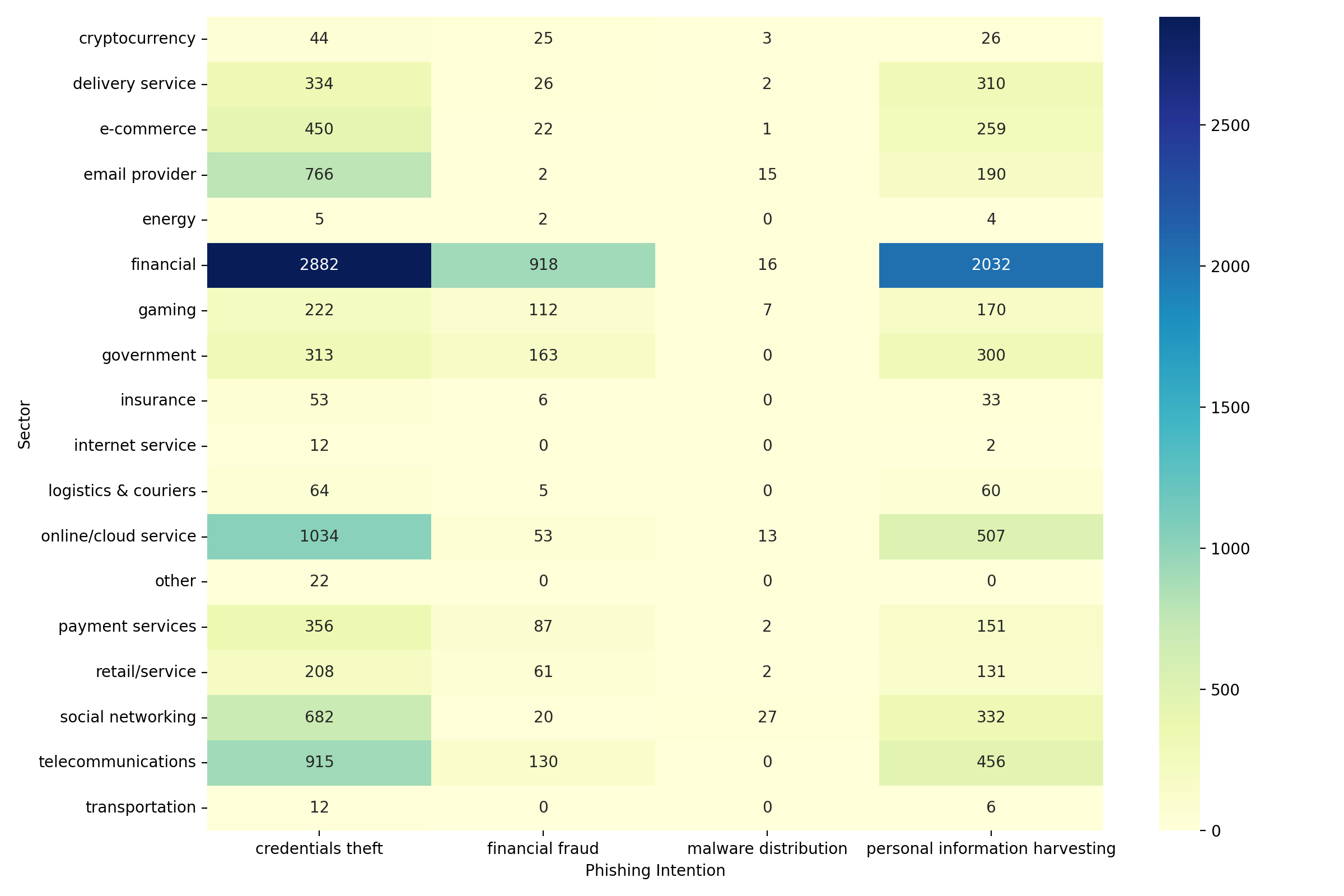}
  \label{intention:sector}
}
\hfill
\subfigure[]{%
  \includegraphics[width=0.50\linewidth]{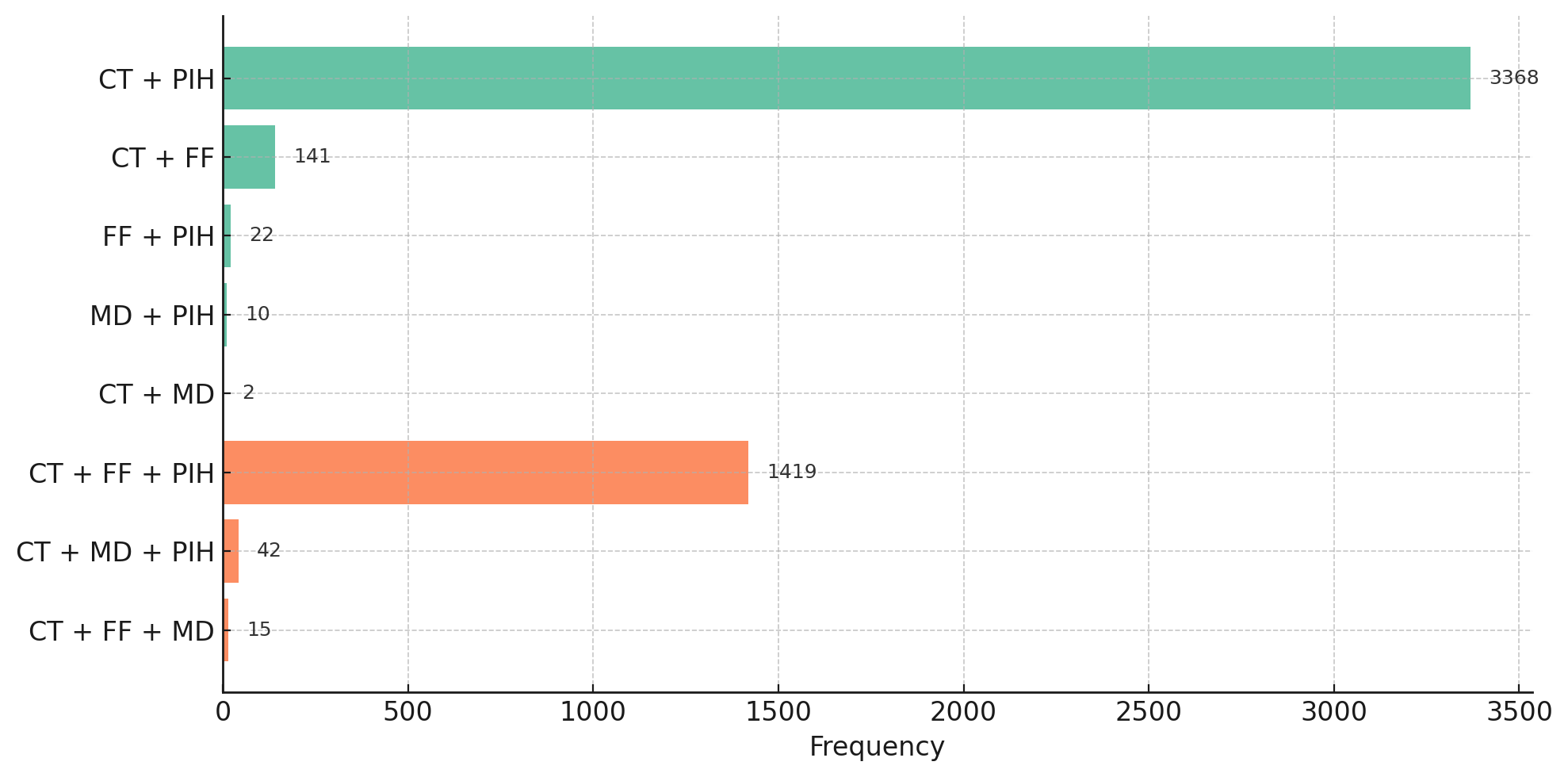}
  \label{fig:intention_combinations}
}

\caption{(a) shows phishing intentions across different sectors; (b) displays frequencies of two- and three-intention combinations. Two-intention combinations are in green; three-intention combinations are in orange. \textbf{Note:} CT = Credentials Theft, PIH = Personal Information Harvesting, FF = Financial Fraud, MD = Malware Distribution.}
\label{fig:intention_combined}
\end{figure}

To understand the phishing intentions in a larger scale, we use our proposed framework with GPT-4o to further evaluate 6K more phishing samples to profile their phishing intentions which results in \textasciitilde 9K samples including our ground truth dataset. A sector–intention frequency matrix was constructed (see Fig.~\ref{intention:sector}). The matrix maps the occurrence of four major phishing intentions: Credentials Theft, Financial Fraud, Malware Distribution, and Personal Information Harvesting—across identified sectors such as financial, e-commerce, telecommunications, and government. The analysis reveals that the financial sector is the most frequently targeted, with 2,882 instances of credentials theft and 2,032 cases of personal information harvesting. Other highly targeted sectors include online/cloud service, email provider, and social networking. Notably, credentials theft and personal information harvesting followed by financial fraud are the dominant intentions across most sectors, while malware distribution remains relatively less frequent.

We further examined records containing multiple phishing intentions to explore the strategic complexity embedded within these campaigns. Fig.~\ref{fig:intention_combinations} shows the frequency and sectoral distribution of co-occurring phishing intentions found in samples with either two or three distinct objectives. Among the two-intention combinations, Credentials Theft + Personal Information Harvesting was by far the most prevalent, appearing in 3,368 instances, with the financial sector being the most frequently targeted (1,185 cases). This combination highlights a dual objective wherein adversaries aim not only to compromise login credentials but also to capture accompanying personal data, thereby increasing the potential for downstream exploitation. The next most common two-intention pair, Credentials Theft + Financial Fraud, appeared in 141 instances, again dominated by attacks on the financial sector (91 cases), suggesting a strong correlation between credential compromise and direct financial gain.

In records exhibiting three distinct phishing intentions, the most frequent combination was Credentials Theft + Financial Fraud + Personal Information Harvesting, found in 1,419 samples, with the financial sector again serving as the primary target (821 occurrences). This triad reflects the layered objectives of modern phishing schemes, where attackers simultaneously seek unauthorized access, financial exploitation, and user profiling. Less frequent but notable three-intention combinations included Credentials Theft + Malware Distribution + Personal Information Harvesting (42 instances, led by the social networking sector) and Credentials Theft + Financial Fraud + Malware Distribution (15 instances, with the gaming sector most affected), demonstrating that more complex phishing strategies often align with the sector-specific threat surface and user value.

The prevalence of such multi-intention combinations likely reflects an adaptive response by attackers to increasingly robust verification mechanisms used by modern systems. As single data points such as passwords or email addresses are often insufficient to gain full access to target systems—particularly those with multi-factor authentication or behavioral risk scoring—phishing campaigns have evolved to collect multiple types of sensitive information concurrently. This multi-vector approach significantly enhances the likelihood of bypassing layered defenses and achieving the attackers' ultimate goals.

\section{Conclusion}\label{sec:conclusion}

This study presents \textit{PhishIntentionLLM}, a novel multi-agent RAG framework for uncovering phishing website intentions through screenshot analysis. While our results demonstrate strong performance and scalability, future work can explore real-time deployment to profile phishing intentions in the wild. Additionally, the proposed approach holds potential for broader application, such as recognizing attacker intentions in phishing emails and other social engineering vectors.

%
%

%
%
%
\bibliographystyle{splncs04}
\bibliography{mybibliography}
%




\end{document}